\documentclass{article}

\usepackage{arxiv}

\usepackage[utf8]{inputenc} 
\usepackage[T1]{fontenc}    
\usepackage{hyperref}       
\usepackage{url}            
\usepackage{booktabs}       
\usepackage{amsfonts}       
\usepackage{nicefrac}       
\usepackage{microtype}      
\usepackage{lipsum}
\usepackage{graphicx}
\graphicspath{ {./images/} }
\usepackage{amsmath}
\usepackage{graphicx}
\usepackage{hyperref}
\usepackage{mathrsfs}
\usepackage{amsfonts}
\usepackage{booktabs} 
\usepackage{caption}  
\usepackage{threeparttable} 
\usepackage{algorithm}
\usepackage{float}
\usepackage{algorithmicx}
\usepackage{algpseudocode}
\usepackage{listings}
\usepackage{enumitem}
\usepackage{chngcntr}
\usepackage{booktabs}
\usepackage{lipsum}
\usepackage{multirow}
\usepackage{subcaption}
\usepackage{csquotes}       
\usepackage{diagbox}
\usepackage{nopageno}
\usepackage{textcomp}

\title{Automatic Prostate Volume Estimation in Transabdominal Ultrasound Images\thanks{This work has been submitted to Radiology: Artificial Intelligence for possible publication. Copyright may be transferred without notice, after which this version may no longer be accessible.}}

\author{Tiziano Natali \thanks{T. Natal and L. Kurucz contributed equally to this work.}\\
Image-Guided Surgery\\Department of Surgical Oncology\\ Netherlands Cancer Institute\\ The Netherlands\\
Department of Nanobiophysics\\ Faculty of Science and Technology\\ University of Twente\\ The Netherlands
\And
Liza M. Kurucz $^\dagger$ \\\
Center for Early Cancer Detection\\  Netherlands Cancer Institute\\ The Netherlands\\
Department of Nanobiophysics\\ Faculty of Science and Technology\\ University of Twente\\ The Netherlands
\And
Matteo Fusaglia\\
Image-Guided Surgery\\ Department of Surgical Oncology\\  Netherlands Cancer Institute\\ The Netherlands
\And
Laura S. Mertens\\
Department of Urology\\  Netherlands Cancer Institute\\ The Netherlands\\
Center for Early Cancer Detection\\ Netherlands Cancer Institute\\ The Netherlands
\And
Theo J.M. Ruers\\
Image-Guided Surgery\\ Department of Surgical Oncology\\ Netherlands Cancer Institute\\ The Netherlands\\
Center for Early Cancer Detection\\  Netherlands Cancer Institute\\ The Netherlands
\And
Pim J. van Leeuwen\\
Department of Urology\\  Netherlands Cancer Institute\\ The Netherlands\\
Center for Early Cancer Detection\\  Netherlands Cancer Institute\\ The Netherlands
\And
Behdad Dashtbozorg\\
Image-Guided Surgery\\ Department of Surgical Oncology\\Netherlands Cancer Institute\\ The Netherlands\\
Center for Early Cancer Detection\\  Netherlands Cancer Institute\\ The Netherlands
}

\begin{document}
\maketitle
\begin{abstract}
\textbf{Introduction:} Prostate cancer is a leading health concern among men, requiring accurate and accessible methods for early detection and risk stratification.
Prostate volume (PV) is a key parameter in multivariate risk stratification for early prostate cancer detection, commonly estimated using transrectal ultrasound (TRUS). 
While TRUS provides precise prostate volume measurements, its invasive nature often compromises patient comfort. 
Transabdominal ultrasound (TAUS) provides a non-invasive alternative but faces challenges such as lower image quality, complex interpretation, and reliance on operator expertise. 
This study introduces a new deep-learning-based framework for automatic PV estimation using TAUS, emphasizing its potential to enable accurate and non-invasive prostate cancer risk stratification.

\textbf{Methods:} A dataset of TAUS videos from 100 individual patients (median 67, 95-percentiles 55 and 81.2) was curated, with manually delineated prostate boundaries and calculated diameters by an expert clinician as ground truth. 
The introduced framework integrates deep-learning models for prostate segmentation in both axial and sagittal planes, automatic prostate diameter estimation, and PV calculation.
Segmentation performance was evaluated using Dice correlation coefficient (\%) and Hausdorff distance (mm).
Framework's volume estimation capabilities were evaluated on volumetric error (mL).

\textbf{Results:} 
The deep learning model trained in the axial plane outperformed the sagittal model in  regards to segmentation performance. 
The axial model achieved an overall Dice score of 0.76 ± 0.16 compared to 0.68 ± 0.21 for the sagittal model, a Dice-MidPlane of 0.91 ± 0.06 versus 0.83 ± 0.10, and a Hausdorff distance of 6.21 ± 4.33 mm compared to 7.93 ± 4.27 mm, respectively.
The framework demonstrates that it can estimate PV from TAUS videos with a mean volumetric error of -5.5 mL (95\% limits of agreement: -13.7 to 13.5 mL), which results in an average relative error between 5 and 15\%. 

\textbf{Conclusion:} The introduced framework for automatic PV estimation from TAUS images, utilizing deep learning models for prostate segmentation, shows promising results. It effectively segments the prostate and estimates its volume, offering potential for reliable, non-invasive risk stratification for early prostate detection.\end{abstract}


\section{Introduction}
Prostate cancer is the second most common form of cancer in adult men, with approximately 1.4 million new cases diagnosed every year (\cite{bray2024global}), with poorer patient prognosis if detected at later stages.
Therefore, it is key to detect prostate cancer at an early stage, when treatment can be less invasive and with higher chance of successful cure.
The current early prostate cancer diagnostic procedures revolve around different nomograms, aimed at quantifying prostate cancer risk based on clinically predictive parameters (\cite{roobol2010risk, Morlacco2021}).

One commonly used indicator for assessing prostate cancer risk is the blood level of Prostate-Specific Antigen (PSA). However, relying solely on PSA levels may not always provide the most accurate way of assessment due to a lack of specificity, since PSA elevation can also occur in benign conditions such as benign prostate enlargement. A more informative measure is  PSA density, which takes into account the size of the prostate. 
By comparing PSA levels  to the prostate’s volume, PSA density offers and helps reduce the risk of  false positives, thus reducing the number of false positive diagnosis in the prostate cancer risk stratification process.

Prostate volume (PV) is typically measured during the initial stage of diagnostic evaluations (\cite{Yusim2020}).
In most cases, the standard clinical practice involves estimating PV based on prostate diameters obtained from transrectal ultrasound (TRUS) exams.
These methods provide valuable insights into prostate size, which are critical for assessing prostate diseases (\cite{PVL2017}).
However, TRUS examinations often come with patient discomfort and sometimes fear due to the invasiveness of rectal examination (\cite{romero2008, walker2005}). 
Additionally, the procedure requires trained clinicians and specialized hardware.
Consequently, TRUS is a sub-optimal method for the estimation of PV in the risk stratification for early prostate cancer detection.

Transabdominal ultrasound (TAUS) is a possible alternative to TRUS, offering a less invasive, easily operable and widely available solution.
TAUS is an affordable imaging modality, that can generate high-contrast images, real-time imaging, and short examination time. 
Its greatest advantage is the ability to enable risk stratification at an early stage, without requiring specialized radiologists or referrals, making it potentially applicable even in primary care settings.
Recent studies (\cite{de2023prostate, Guo2023}) also reported TAUS as a valid alternative imaging method for the estimation of PV.
There the authors showed that TAUS can provide reasonable estimates of prostate volume with reduced patient discomfort, as it does not require a rectal examination.
Furthermore, examination time can be decreased and examinations can be performed by clinicians with less specialized training, making it a more practical and accessible solution for routine PV estimation in prostate cancer diagnostics.

Despite its advantages, using TAUS for PV estimation comes with several challenges. One key issue is the limited image resolution and low signal-to-noise ratio compared to TRUS, as the distance between the ultrasound probe and the prostate is greater. 
This can lead to a more difficult image interpretation, less prostate visibility, and less precise measurements.
Additionally, TAUS is more susceptible to interference from factors such as abdominal fat, bladder fullness, and bowel gas, all of which can compromise the user's interpretation and quality of the images obtained. 
Additionally, the acquisition procedure requires the clinician to identify the imaging plane that has the largest prostate section and then measure the prostate dimensions. 
This process is highly dependent on the operator and prone to errors.

In recent years, AI solutions have demonstrated significant potential in the analysis of US videos. 
These technologies can automatically interpret US frames and provide instant diagnostic aid directly to clinicians during examinations, possibly enhancing the accuracy and efficiency of the diagnostic process.
As described by \cite{kurucz2024advances} these methods can achieve promising results in the detection and segmentation of the prostate from TRUS frames (\cite{anas2018deep, wang2024lightcm, beitone2022, Xu2022}).
In the study by \cite{beitone2022}, the authors implemented a framework using three models to segment the prostate along three imaging planes from TRUS frames, showcasing that the integration of data from different planes improves segmentation performances.
Although this work achieved promising segmentation results, it is not possible to obtain US videos along three imaging planes when examining trans-abdominally.
Although much progress has been made on TRUS videos, little translation has been done to TAUS videos.
\cite{natali2024} proposed Nano-DeTr, a model for the real-time detection of prostate in TAUS videos.
Here they managed to implement an accurate detection network with a relatively small dataset consisting of 55 US videos, suggesting the feasibility of automatic prostate segmentation with an AI approach.

In this study, we present for the first time, a framework for the automatic estimation of prostate volume from TAUS acquisitions. 
The framework is meant to improve the prostate cancer diagnostic process, decreasing the number of referrals to specialists, and reducing the hospital workload by providing an accurate tool for prostate volume estimation from TAUS videos to inexperienced US operators.

Overall, the primary contributions can be summarized as follows:
\begin{itemize}
    \item Collection of a dataset consisting of TAUS scans along the sagittal and axial imaging planes from a population of 100 patients (median age 67, 95-percentiles 55 and 81.2).
    \item Development of deep learning models for automatic prostate segmentation along the two imaging planes.
    \item Introduction and evaluation of a framework for the automatic prostate volume estimation based on prostate segmentation masks in TAUS video sweeps.
\end{itemize}

\section{Methods}
\begin{figure}[!h]
    \centering
    \includegraphics[width=0.9\linewidth]{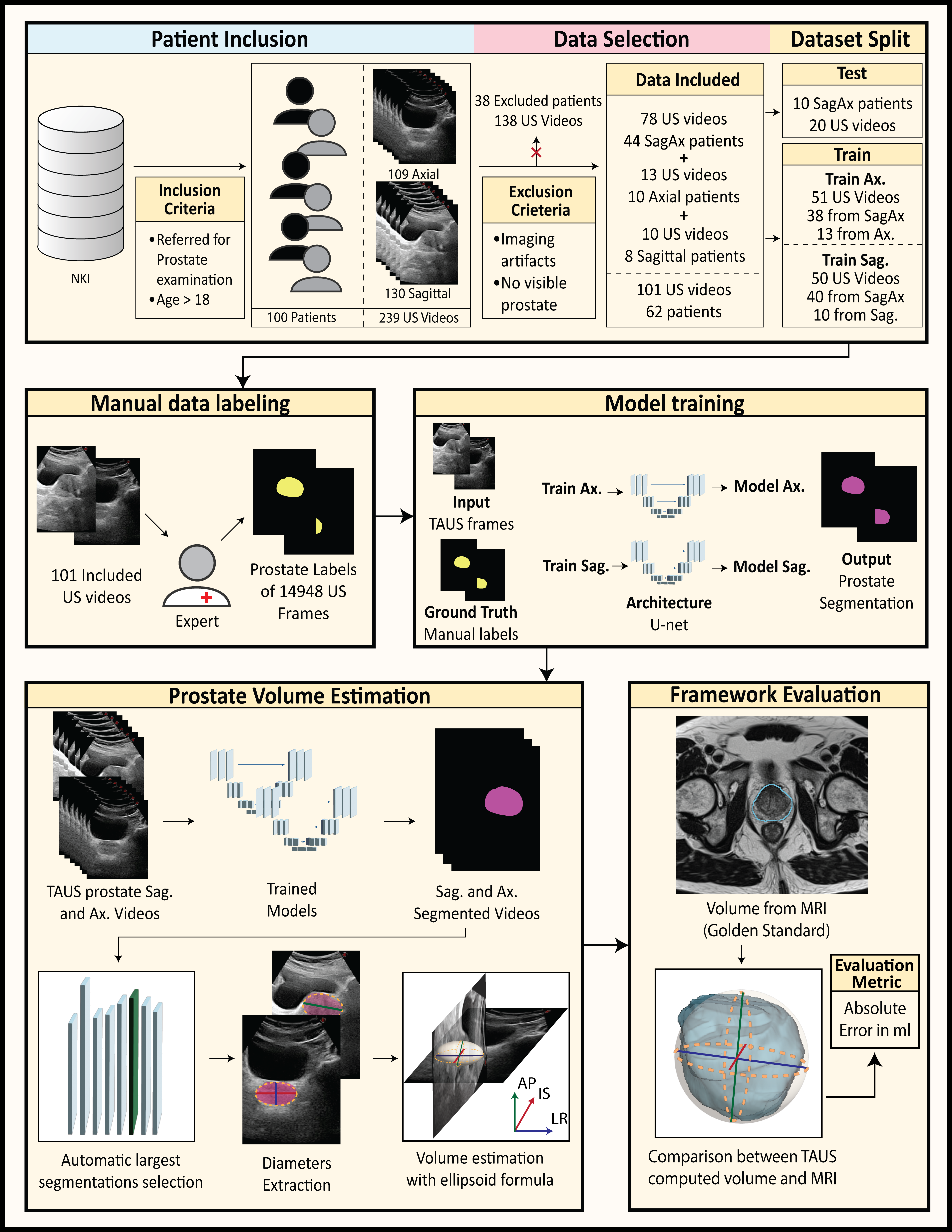}
    \captionsetup{font=small}
    \caption{ Overview of the study. A dataset of 100 patients has been acquired between August 2023 and February 2024. Prostates of all patients were scanned with a TAUS transducer, and at least one US video from the axial and one along the sagittal imaging plane was acquired. As a result of a quality check (improper transducer handling or technical issues with the scanner), US videos from 38 patients have been excluded. US videos from the remaining patients were split into training and testing sets. Two deep-learning models, one per imaging plane, were trained to automatically segment the prostate from TAUS frames. Prostate diameters were extracted and used to calculate the prostate volume using the ellipsoid formula and then compared to their respective reference-standard values (MRI).}
    \label{fig:project_overview}
\end{figure}

This study presents a framework for the automatic estimation of prostate volume from TAUS frames, which is presented by Fig.\ref{fig:project_overview}.
A dataset including 100 individual patients was created during the study, and, for each patient, prostate TAUS videos along axial and sagittal imaging planes were acquired.
The presented framework uses two deep-learning models to segment the prostate, trained on ground truth segmentation masks provided by a clinician.
The segmentation masks are then used to automatically extract three prostate diameters, which are finally used in an ellipsoid formula to compute the prostate volume.

\subsection{Data Acquisition}
This prospective feasibility study adhered to the principles of the Declaration of Helsinki and received approval from the institutional review board (IRBd22-319) of the Netherlands Cancer Institute-Antoni van Leeuwenhoek Hospital in Amsterdam, The Netherlands.
Patients undergoing evaluation for early prostate cancer detection with a scheduled MRI scan at the NKI Center for Early Cancer Detection were included in the study between August 2023 and February 2024.
After oral and written introduction to the study, a written informed consent form was obtained from each participant.
In order to prove the feasibility of the proposed methods, the study was designed to acquire data from 100 individual patients included sequentially in the study.
Immediately following the MRI scan, subjects underwent transabdominal ultrasound imaging, where the TAUS operator was blind to the MRI results.
At least one ultrasound video was acquired in the sagittal plane and one in the axial plane, with multiple videos captured in some cases.

Ultrasound videos (sweeps) were used instead of single snapshots (frames) images to enhance the accuracy and reliability of prostate volume estimation. Videos provide a continuous sequence of frames, capturing a broader range of anatomical details and reducing the limitations associated with single-image acquisition. 
This approach allows for improved segmentation by analyzing multiple perspectives and reduces operator dependency by offering more frames to select from during data processing. 
Additionally, the use of videos enlarges the dataset for training deep learning algorithms, enabling the model to learn temporal patterns and variations in anatomy.
The US videos were performed so that the prostate was acquired entirely, from apex to base in the axial plane and from left to right in the sagittal plane.
The data acquired consists of a total of 102 sagittal and 102 axial TAUS videos, each containing 150 to 200 US frames.
All videos were acquired using the same BK3000 US scanner (BK-Medical, Herlev, Denmark) coupled with a 6C2 (9040) curved array transducer (Frequency Range 6-2 MHz) using the preset for bladder scanning.
The videos were acquired by an urologist or nurse practitioners who were initially trained by the urologist.
All data was de-identified with a re-identification key.

\subsection{Dataset Partitioning}
\label{sec:partitioning}

Data acquired from the 100 individual patients have been first manually checked to ensure image quality, excluding US videos presenting large imaging artifacts, or echoic shadows due to improper handling, or improper imaging settings.
Given the US video exclusion criteria, included patients could result in having videos, included in the final selection, only from one of the imaging planes ($Patients_{Ax}$ and $Patients_{Sag}$) or from both ($Patients_{SagAx}$).

The test set (referenced as $TestSet$) was formed by selecting the US videos from 10 individual patients in the $Patients_{SagAx}$ set, which is 10\% of the patient population. 
The rest of the selected US videos was used to form the training sets ($TrainSet_{Ax}$, $TrainSet_{Sag}$) of the segmentation models ($Model_{Ax}$, $Model_{Sag}$) along the axial and sagittal imaging planes, respectively.
Furthermore, train sets along both imaging planes were divided at patient level into 4 subsets of similar sizes.
The subsets were necessary to perform 4-fold cross-validation of the trained segmentation models.

In order to ensure data variability, frames included in both training sets were sampled from US videos by extracting every fifth frame.
All frames from the videos included in the test set were used for evaluation, in order to better represent the final use-case of the framework.

\subsection{Ground truth delineation}
\label{sec:gt}

\begin{figure}[!h]
    \centering
    \includegraphics[width=0.7\linewidth]{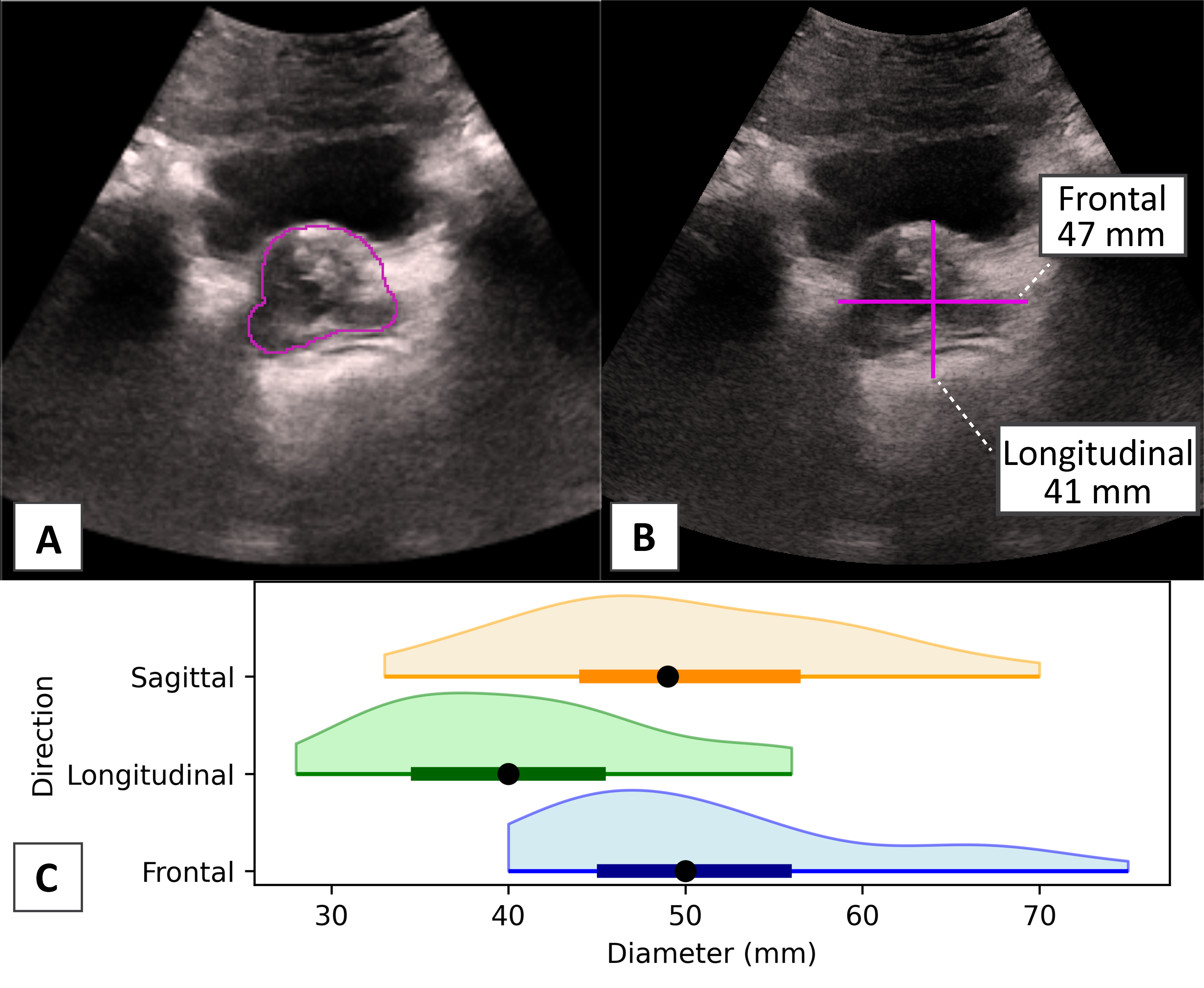}
    \captionsetup{font=small}
    \caption{Example of manual delineation process. For each frame in the US videos, a clinician provided a boundary annotation of the prostate. At the same time, prostate diameters were measured in the TAUS videos. A) Ground truth segmentation mask based on boundary annotation. B) Extracted diameters, in this example a longitudinal and a frontal diameter. C) Distributions of the manually delineated diameters. Longitudinal and frontal diameters are extracted from the axial US videos and the sagittal diameter from sagittal ones.}
    \label{fig:prostate_fig1}
\end{figure}

Manual prostate boundary delineations were performed by an expert for all US frames in the training and test sets as described in Tab.\ref{tab:datasets}.
The manual prostate delineations were converted into binary masks which were used as ground truth for training and testing the two segmentation models.
Additionally, the clinician measured the prostate diameters in each US video.
Fig.\ref{fig:prostate_fig1}A presents an example of the prostate delineations, while an example of the diameters manually extracted by the clinician is pictured in Fig.\ref{fig:prostate_fig1}B.
Labelling procedures have been performed in 3D Slicer 5.6.2 \cite{pieper20043d}.
These measurements were taken from the frame where the prostate appeared largest, identified by visual inspection of the entire video sweep, similar to the process used during the TRUS examination.
The manually obtained diameters were used as ground truth in the evaluation of the prostate volume estimation algorithm of the entire framework.
During the labeling procedures, the clinician reviewed all the frames in the video to ensure high accuracy in the delineation of all diameters and prostate boundaries.

For each patient in the $TestSet_{volume}$, the prostate volume estimated from the MRI acquisition (computed from three manually assigned diameters as the standard of care procedure) was also extracted. These values were used as the reference-standard for prostate volume for the performance evaluation of the entire framework.

Finally, to evaluate intraobserver agreement,  three extra sets of prostate delineations for the US frames in the $TestSet_{volume}$ were provided by three medical technicians, all of whom were initially instructed on the procedure by a urologist.

\subsection{Deep-Learning Model}

Two separate deep-learning models were employed for prostate segmentation in the sagittal and axial imaging planes due to the unique characteristics of each plane. 
Each plane offers distinct anatomical perspectives, capturing features that may not be apparent in the other. 
In the work from \cite{beitone2022}, the authors proposed a framework for the segmentation of prostate TRUS, where they implemented segmentation models for each imaging plane, showing that improved results developing dedicated models for each plane allows to achieve improved segmentation results. 
The development of models tailored to each imaging plane enables their optimization over imaging characteristics specific to each plane, thus enhancing their segmentation accuracy.
This approach also facilitates the validation of segmentation results, where discrepancies between the planes can highlight potential inaccuracies. 
Additionally, separate models address variations in image quality and noise, potentially improving the segmentation accuracy. 
Ultimately, both models leverage unique information, ensuring a robust and comprehensive segmentation framework. 
To implement this approach, one model was trained to segment the prostate in axial TAUS frames ($Model_{Ax}$) using the $TrainSet_{Ax}$, while the second was trained to segment from sagittal TAUS frames ($Model_{Sag}$) using the $TrainSet_{Sag}$.

Models input a single US frame and output a binary mask with the delineation of the prostate.
Both models were trained using the nnU-Net model (\cite{isensee2021nnu}) in its standard configuration, which implements Dice + Cross-entropy as the loss function, training of 1000 epochs with  Stochastic gradient descent (SGD) optimizer, and a learning rate of 0.0001 with logarithmic decay. The nnU-Net model was selected due to its proven adaptability and high performance across various medical image segmentation tasks.
Furthermore, its dynamic configuration of the hyperparameters, which are tailored to dataset-specific characteristics, minimizes the need for extensive manual tuning.
Hence, this setup will make for an easily reproducible baseline that can be implemented in future works in this field, either for development or comparison purposes.
The models have been trained using Python 3.10, and Pytorch 2.3 on a machine equipped with an Nvidia Geforce\textsuperscript{\textregistered} 1080 Ti with CUDA 11.8 and an Intel Xeon\textsuperscript{\textregistered} W-1250P (12M Cache, 4.10 GHz).

\subsection{Prostate Diameters Extraction}

In order to compute the prostate volume from the predicted prostate segmentation masks, the ellipsoid formula in \autoref{eq:ellipsoid} was adopted, since it is the most commonly used method for calculating prostate volume from MRI as well as in transrectal ultrasound images (\cite{Guo2023, de2023prostate}).

\begin{equation}
PV_{Ellipsoid} = \quad D_{frontal} \quad \times \quad D_{longitudinal}\quad \times \quad D_{sagittal} \quad \times \quad \pi / 6    
\label{eq:ellipsoid}
\end{equation}

The diameter extraction procedure implemented the following steps: (i) all images in acquisition were input to the respective segmentation model; (ii) the generated segmentation masks were then analyzed to find the prostate mask with the largest area (i.e., number of pixels); (iii) the previous step was repeated twice per patient, once for an axial acquisition and once for the sagittal; (iv) an ellipsoid was fit through the perimeter of the selected segmentation masks; (v) two diameters ($D_{frontal}$ and $D_{longitudinal}$) were extracted from the ellipsoid fit in the largest prostate mask of the axial acquisition, and a single diameter ($D_{sagittal}$) from the sagittal; (vi) the prostate volume is computed using the \autoref{eq:ellipsoid}.

\subsection{Evaluation Metrics and Experimental Setup}

The prostate segmentation accuracy of the trained models was first evaluated. 
The segmentation masks in the test set obtained with $Model_{Ax}$ and $Model_{Sag}$ were compared at pixel level against the ground truth delineation (see Sec.\ref{sec:gt}), computing Dice, Dice Mid-Plane and Hausdorff Distance Mid-Plane (HD Mid-Plane).
The latter was adopted to understand the diameter estimation error achieved with this approach (i.e., diameter extraction from segmentation masks) in relation to the distribution of manually extracted diameters (as introduced in Fig.\ref{fig:prostate_fig1}).
Additionally, a four-fold cross-validation was performed to better evaluate the consistency of $Model_{Sag}$ and $Model_{Ax}$ on a virtually larger number of patients, thus better assessing their stability and generalization properties through the different folds and patients.

Inter-observer variability of the segmentation was computed on the same three metrics by comparing the three extra sets of delineations of the test set against the reference-standard provided by the clinician.
Results are reported as the average over the three comparisons.

The entire workflow was then evaluated on the 20 acquisitions from ten patients in the $TestSet$ (\autoref{sec:partitioning}).
Although TRUS is the most common imaging modality adopted in the risk stratification for early prostate cancer detection, because of its accessibility, speed, and inexpensive nature, MRI has been shown in previous studies to be the most accurate imaging method when assessing prostate volume (\cite{Youn2023}).
Therefore, in this study, MRI was used as reference-standard in the evaluation of the volumetric estimation capabilities of the proposed framework.
The volumes estimated with the diameters automatically extracted from the TAUS images were compared to the respective reference-standard volumes computed by a radiologist on the MRI scans, as per the study design.
The evaluation metric used in this test was volume difference, calculated in $mL$ and presented in a Bland Almand plot, where the error is computed by subtracting the predicted volumes from their respective reference-standard values.

\section{Results}
\subsection{General Data}

\begin{table}[b!]
\centering
\captionsetup{font=small}
\caption{Number of patients, US videos, and frames per direction included in the dataset. As mentioned in \autoref{fig:project_overview}, videos from both directions were selected for all the patients in the test set, and for part of those in the training set. Videos from one direction only were selected for the rest of the included patients. Every fifth frame from the US videos in the training set was included in order to ensure data variability when training the DL models. On the other hand, all frames in the videos of the test set were included.}
\begin{tabular}{lcclccc}
\hline
\textbf{Split}                  & \multicolumn{1}{l}{\textbf{Patients}} & \multicolumn{1}{l}{\textbf{US Videos}} & \textbf{Direction}    & \multicolumn{1}{l}{\textbf{Patients}} & \multicolumn{1}{l}{\textbf{US Videos}} & \multicolumn{1}{l}{\textbf{US Frames}}\\ \hline\hline
\multirow{2}{*}{\textbf{Train}} & \multirow{2}{*}{52}          & \multirow{2}{*}{101}           & Axial           &  44                          & 51                            &  1516                        \\
                       &                              &                               & Sagittal          &  42                          & 50                            &  1420                        \\ \hline
\multirow{2}{*}{\textbf{Test}}  & \multirow{2}{*}{10}          & \multirow{2}{*}{20}           & Axial           &  10                          & 10                            &  1561                        \\
                       &                              &                               & Sagittal          &  10                          & 10                            &  1627                        \\ \hline
\end{tabular}

\label{tab:datasets}
\end{table}

Age distribution of the 100 included patients resulted to have a median of 67 years with 95\% confidence intervals of 55 and $81.2$.
\autoref{fig:prostate_fig1}c, presents the distribution of the diameters manually extracted from the TAUS videos of the 100 patients included in the study.

The dataset quality check (see \autoref{sec:partitioning}) resulted in an exclusion of 138 videos from 38 patients and a selection of 101 US videos from 62 patients.
The majority of the excluded US videos were performed during the first phases of the study, hinting that the operators required experience to improve their examination skills.
For 44 of the 62 selected patients, US videos in both imaging planes were available (referenced as patients SagAx), while for 14 only the axial US videos were available (patients Ax), and for the remaining 9 only the sagittal ones (patients Sag).

\autoref{tab:datasets} summarizes the number of US videos and patients in each of the sets described in \autoref{sec:partitioning}.
$Model_{Ax}$ was trained on 1516 frames that were sampled from 51 videos of 44 individual patients, videos of which were 38 from the $Patients_{SagAx}$ and 13 from $Patients_{Ax}$ sets.
$Model_{Sag}$ was trained on 1420 frames that were sampled from 50 videos of 42 individual patients, of which 40 from the $Patients_{SagAx}$ and 10 from $Patients_{Sag}$ sets.
The test set included 1561 axial and 1627 sagittal US frames, from 10 and 10 US videos from 10 individual patients that were part of the $Patients_{SagAx}$ set.

\subsection{Segmentation Models Evaluation}
\begin{figure}[!ht!]
    \centering
    \includegraphics[width=\textwidth]{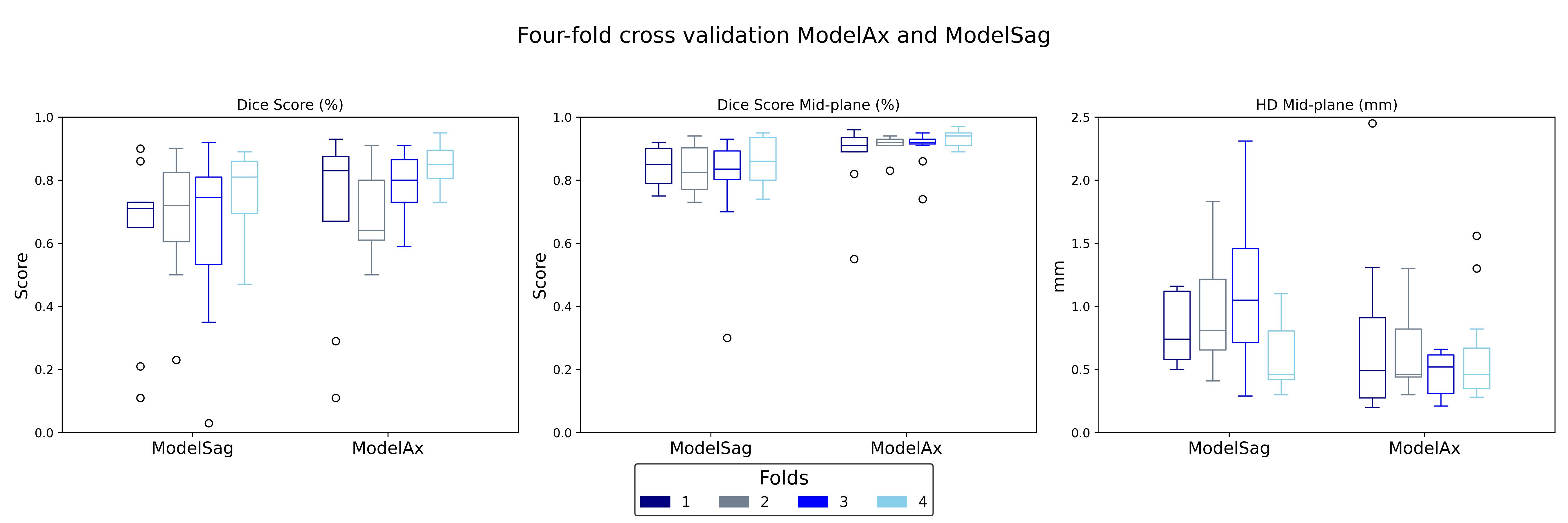}
    \captionsetup{font=small}
    \caption[The evaluation of $Model_{Ax}$ and $Model_{Sag}$ in a four-fold cross-validation.]{Model performance summarized in boxplots for each fold in the cross-validation of $Model_{Ax}$ and $Model_{Sag}$ on the $TrainSet_{Ax}$ and $TrainSet_{Sag}$ splits, respectively. For each fold, Dice, Dice Mid-Plane, and HD Mid-Plane are reported.}
    \label{fig:cross_val}
\end{figure}

The boxplots in \autoref{fig:cross_val} summarize the results for Dice, Dice Mid-plane, and HD of the four-fold cross-validation for $Model_{Ax}$ and $Model_{Sag}$. 
The two models behaved similarly, where both scored lower values for Dice with larger error margins, and higher values for Dice Mid-plane with smaller error margins.
The results show a Dice score difference of 0.16 between the best and worst-performing folds (0.69 and 0.85, respectively) in the case of $Model_{Ax}$, and 0.14 (0.62 and 0.76) for $Model_{Sag}$.
Both models consistently achieved results for the Dice Mid-plane higher than 0.91 for $Model_{Ax}$, and 0.83 for $Model_{Sag}$, which is the most relevant performance metric for the proposed framework.
Although the scores for HD present relatively high standard deviations, the results show that both models could segment the prostate with mean HD lower than $0.62 \pm 0.43$ and $0.89 \pm 0.42$ mm, respectively for $Model_{Ax}$ and $Model_{Sag}$.
Thus, the shapes of the automatic prostate segmentation masks are mostly similar to their respective ground truth.
If the computed values of HD are compared to the medians of the those manually extracted by the expert in \autoref{fig:prostate_fig1}C (48.9, 40.1, 50.0 mm for sagittal, longitudinal, and frontal, respectively), it follows that the error in the volume estimation induced by automatically segmenting the prostate with the trained models is only marginal.

The results present an overall improved performance of $Model_{Ax}$ over $Model_{Sag}$ across all metrics, indicating that the segmentation of the prostate in the axial US plane is less challenging compared to those in sagittal.

\autoref{tab:segmentation_results} further summarizes the four-fold cross-validation of $Model_{Ax}$ and $Model_{Sag}$, presenting an overview of the results in \autoref{fig:cross_val}, by averaging the values for the three metrics over the four folds.
The scores in the table show a clearly improved performance of $Model_{Ax}$ over $Model_{Sag}$ across all metrics, where the latter achieved 8\% lower Dice (0.68 vs 0.76) and Dice-MidPlane (0.83 vs 0.91) and a higher Hausdorff distance by 2.68 mm (8.91 vs 6.23).
Furthermore, the table compares the results achieved by the trained models against the inter-observer variability.
The performance of the models is comparable to the inter-observer variability computed on the US frames in our test set, whereby the models scored: lower scores by 0.01 and 0.02 for Dice; a Dice Mid-Plane higher by 0.01 on the axial direction and lower by 0.01 along the sagittal; a higher HD Mid-Plane by 0.13 and 0.17 mm on the axial and sagittal direction, respectively.
When HD Mid-Plane results are compared to the prostate diameters size distributions, it can be seen that the models are achieving relative errors between 5\% and 15\%.

\begin{table}[b!]    
    \centering
    \captionsetup{font=small}
    \caption[Performance of developed models for prostate segmentation on transabdominal ultrasound]{Average Dice, DSC mid-plane, and HD Mid-Plane of the results achieved in the by $Model_{Ax}$ and $Model_{Sag}$ for the prostate segmentation in the US frames in the Test set compared to the inter-observer performance. Metrics are computed per image at pixel level and here are reported the averages over the images in the test set $\pm$ standard deviation. 
    }
    \begin{tabular}{lccc}
    \hline
        \textbf{Model}              & \textbf{Dice}    & \textbf{Dice Mid-Plane} & \textbf{HD Mid-plane (mm)}\\ \hline\hline
        $Model_{Ax}$            & $0.76 \pm 0.16$  & $0.91 \pm 0.06$        & $6.21 \pm 4.33$           \\  
        $Model_{Sag}$           & $0.68 \pm 0.21$  & $0.83 \pm 0.10$        & $7.93 \pm 4.27$           \\ \hline
        $\text{Inter-observer}_{Ax}$  & $0.77 \pm 0.23$  & $0.90 \pm 0.12$        & $6.08 \pm 3.79$           \\
        $\text{Inter-observer}_{Sag}$ & $0.70 \pm 0.28$  & $0.84 \pm 0.17$        & $7.76 \pm 4.11$           \\ \hline
    \end{tabular}
    \vspace{0.5cm}
    \label{tab:segmentation_results}
\end{table}

\begin{figure}[t!]
\centering
\includegraphics[width=\textwidth]{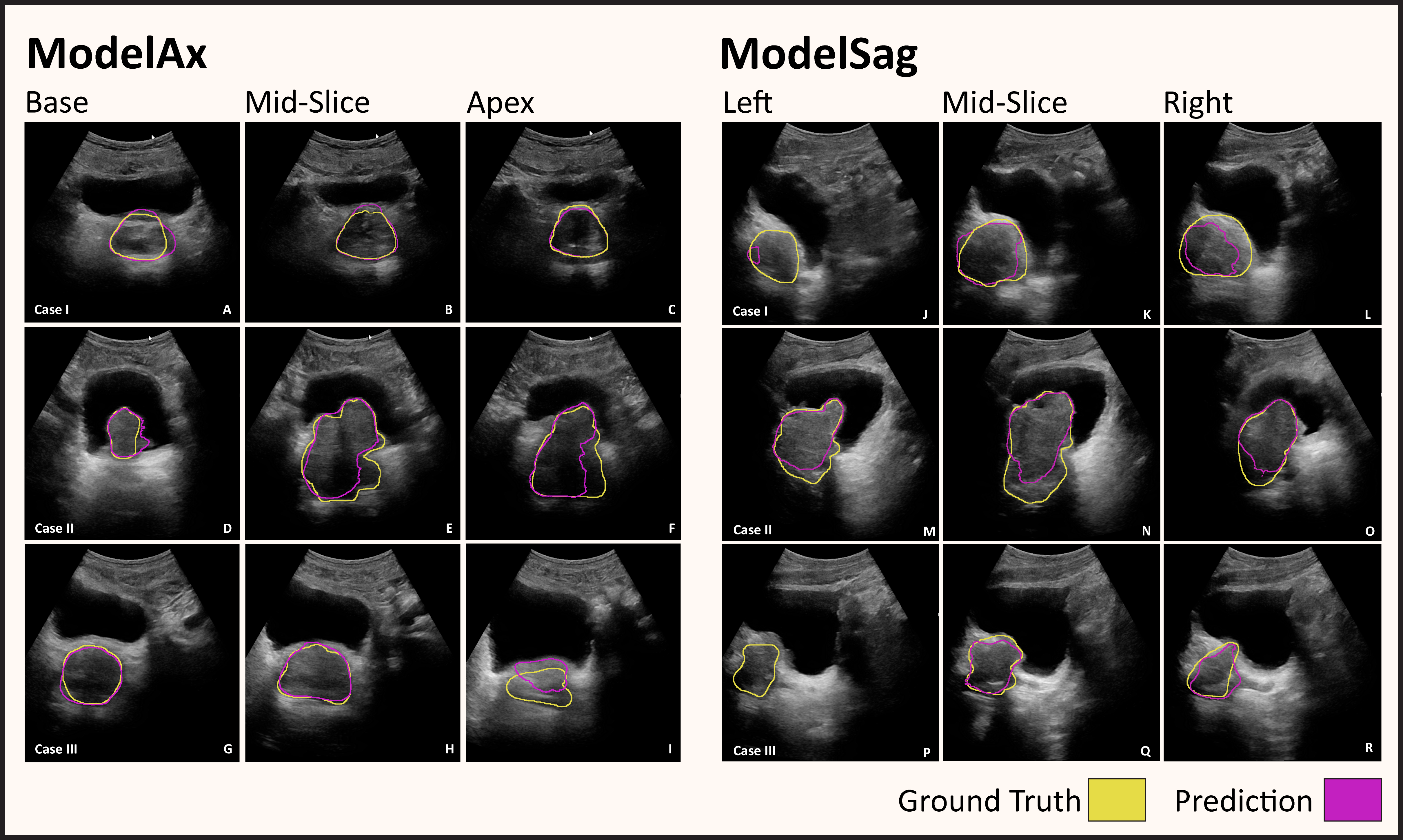}
\captionsetup{font=small}
\caption[Examples of prostate segmentations on axial TAUS images.]{Qualitative analysis of the results from the models for automatic prostate segmentation.
Segmentation results from US videos along both imaging planes from 3 patients.
For each axial sweep, frames from the base, mid-plane, and apex are shown and for each sagittal sweep, frames from left, center, and right are shown.
Both models achieved good segmentation performances on the mid-plane frames.
$Model_{Ax}$ shows improved performances over $Model_{Sag}$, especially when segmenting from US frames of the prostate extremities.
}
\label{fig:qualitative-sag-ax}
\end{figure}

\autoref{fig:qualitative-sag-ax} demonstrates a qualitative overview of the segmentation capabilities of the presented models for prostate segmentation.
Frames of the US videos from three patients along both imaging directions are presented. Prostate base, apex, and mid-plane (i.e., the slice used for diameter extraction) are pictured from left to right for $Model_{Ax}$ and left, mid-plane, and right for $Model_{Sag}$.
From the qualitative assessment of the presented cases, it appears that $Model_{Sag}$ could segment with slightly lower accuracy than $Model_{Ax}$, indicating that prostate segmentation on the sagittal plane is more challenging, as it already appeared in the previous results presented here.
The prostate in the images acquired along sagittal imaging plane is more easily affected by echoic shadows due to the pelvic bone.
As a consequence, the organ visibility is often reduced, possibly inducing lower performances in the segmentation models.
Additionally, the segmentations in the images present a similar trend from the results in \autoref{fig:cross_val} and \autoref{tab:segmentation_results}:
the models segment more accurately the prostate in the mid-plane, with lower performances on US frames from the base and apex of the prostate (most clearly visible in the frames from Case III).

\subsection{Framework Evaluation}

\begin{figure}[!h]
\centering
\includegraphics[width=1\columnwidth]{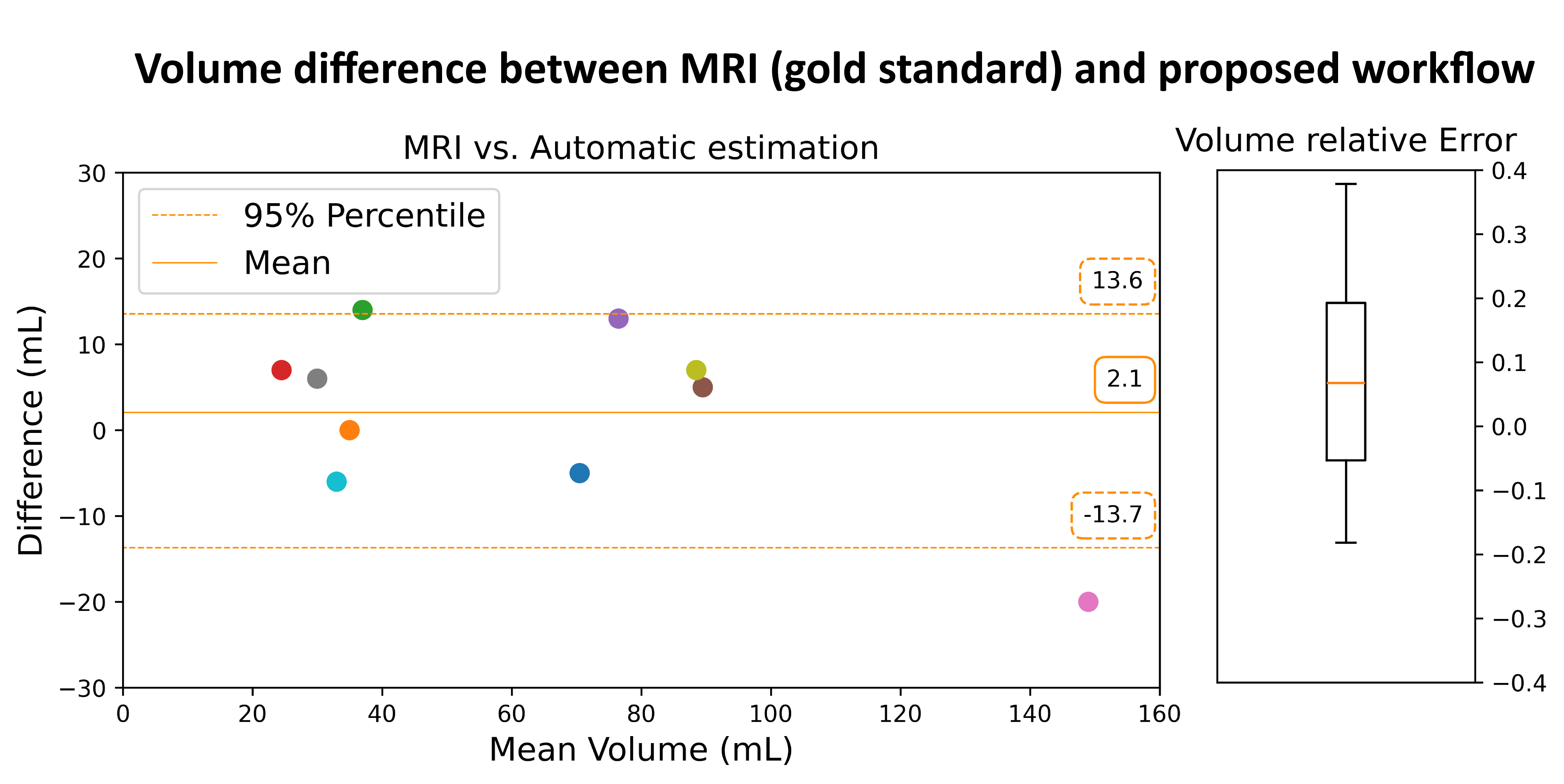}
\captionsetup{font=small}
\caption[Bland Altman plot: the comparison between predicted PV measurements and reference volume measurements based on MRI.]{\textbf{A:} Bland Altman plot comparing the predicted prostate volume (PV) measurements and reference-standard volume measurements based on MRI, on 10 patients in the Test set (\autoref{sec:partitioning}). \textbf{B:}  Distribution of the volume estimation error on the test set (\autoref{sec:partitioning}) relative to prostate dimension computed as average between reference-standard and estimated.}
\label{fig:final_ba_plot}
\end{figure}

\autoref{fig:final_ba_plot} summarizes the results of the proposed framework on the test set ($TestSet_{vol}$).
The estimated prostate volumes with the presented framework are compared to the reference-standard volumes based on MRI.
The left Bland Altman plot shows the difference between reference-standard and estimated volumes against the average of the two measures.
On the right, the boxplot presents the distribution of the volume difference relative to the size of the prostate, where the size is computed as the average between reference-standard and estimated.
The mean error (bias) in the left chart is higher than zero, which would indicate that the proposed workflow tends to underestimate the prostate volume when compared to MRI.
However, it is also visible that in some cases the proposed workflow overestimates the prostate volume, thus resulting in a non-clear over/under-estimation trend.
Volume relative error also shows a trend in underestimation with relatively low dispersion from our proposed framework.
Values revolve around a median of 0.08 and mostly fall between 95-percentile interval of agreement (-0.03 and 0.19), thus suggesting that the automatic prostate volume estimation method maintains consistent performance across different volumes.

\section{Conclusions}
In this study, we explored the implementation of a framework for the automatic estimation of prostate volume from TAUS images and assessed its feasibility, leveraging a deep-learning model for the segmentation of the prostate from TAUS frames and automatic diameter extraction.

Results show that the implemented segmentation models can accurately predict the location of the prostate in an image.
The models demonstrated higher performances when analyzing the largest sections of the prostate, which is the most critical location given that those are the segmentations used in the following steps of the framework.
Computed HD values show that the trained models can commit relative errors that are in the 5\% to 15\% range to the prostate diameter distributions calculated from our population.
$Model_{Ax}$ showed higher performances over $Model_{Sag}$, indicating that prostate segmentation from TAUS frames acquired along the sagittal imaging plane is more challenging.
The reduced visibility of the prostate caused by the echoic shadows in the US videos along the sagittal imaging plane presumably influenced the performances of $Model_{Sag}$.
We think that the implementation of a data augmentation similar to that proposed by \cite{Xu2022} might improve the robustness and performances of $Model_{Sag}$.
This, together with the higher number of discarded US videos along the sagittal imaging plane due to poor quality, suggests that it is more challenging to visualize the prostate in this orientation, thus obtaining TAUS videos of acceptable quality.
When evaluating the entire framework on prostate volume estimation, it achieved values with a mean error of 5 $mL$ within -18, 14 $mL$ 95-percentiles.

While no studies in the literature have reported prostate segmentation performance from TAUS images, our results can still be compared to similar automated approaches on TRUS or to manual estimations of prostate volume from TAUS images.
\cite{karimi2019}, in their study on TRUS videos, achieved automatic prostate segmentation scores of 0.94 on the prostate mid-plane, also showing slightly lower performances on apex and base (0.93 and 0.91, respectively). 
\cite{Guo2023} in their study they compared, on a larger patient population, values from MRI volume estimations to those computed from manually extracted diameters from TAUS frames, achieving an error of 4.1 $mL$ within 95-percentiles of -17.3 and 25.6 $mL$, which are similar to what was also found in this study on a fully automatic estimation.
\cite{de2023prostate}, in their similar study showed that volumes computed from manually extracted diameters from TAUS images achieved an overestimation error of 9.9 $mL$ with 95-percentiles of -5.6 and 13.8, which consists in a slightly larger bias but more consistent error.
This suggests that the errors observed by the proposed framework and presented in our results, are mostly due to the use of different imaging modalities and by limitations of the ellipsoid formula, as reported by \cite{rodriguez2008prostate}, rather than the algorithms employed for automating the volume estimation process.
Therefore, we suggest conducting studies aimed at understanding whether the replacement of the ellipsoid formula with a different volume estimation method would lead to an overall improved accuracy of the framework.
We hypotesize that by leveraging tomographic reconstruction technologies like the one proposed by \cite{prevost20183d}, it will be possible to generate volumetric reconstructions of the prostate from the TAUS videos to more accurately estimate prostate volume.




The US frames included in this study were all acquired using the same ultrasound machine and ultrasound transducer.
A larger study including US frames acquired with different probes from several manufacturers would result in an understanding of the level of generalization of the proposed framework.
The initial quality check on the included patients resulted in an exclusion of 103 US videos due to low quality.
This is related to the fact that some of the users who acquired the videos, had to gain experience in acquiring these videos and by the challenge in avoiding echoic shadows when imaging along the sagittal plane.
Therefore, we suggest future studies focusing on prostate TAUS images to integrate algorithms for automatic prostate detection.
Algorithms like the one proposed by \cite{natali2024} are meant to ease the readability of TAUS images for inexperienced users, thus reducing exclusion rates due to improper examination procedures without putting extra burdens on more expert operators (i.e., urologists, radiologists and trained nurse practitioners).
Moreover, the high exclusion rate of the US videos limited the test set to only 20 videos from 10 patients.
While this test set still made possible the assessment and feasibility of the proposed method, it will still be necessary to expand the test set with a larger and more diverse patient population in order to better evaluating the actual capabilities of the proposed framework.

The proposed framework for automatic prostate volume estimation from TAUS videos scored an average overestimation over MRI-computed volumes of 5 mL, showing that, when TAUS videos are properly acquired, the framework is well-suited for prostate volume estimation.

\bibliographystyle{unsrt}  
\bibliography{references}  






\end{document}